\documentclass[11pt]{article}
\usepackage[cp437de]{inputenc}
\usepackage{epsfig}
\usepackage[intlimits]{amsmath}
\usepackage{latexsym}
\usepackage{amsfonts}

\newcommand{\mS}{\mathcal{S}}

\newcommand{\D}{\mathcal{D}}

\newcommand{\C}{\mathbb{C}}
\newcommand{\Z}{\mathbb{Z}}

\DeclareMathOperator{\sgn}{sgn}
\DeclareMathOperator{\Ei}{Ei}
 
\newcommand{\reseteqn}{\setcounter{equation}{0}}
\newcommand{\mysection}{\reseteqn\section}

\begin{document}


\begin{titlepage}
\noindent
December, 1996 \hfill IASSNS-HEP-96/130 \\
revised April, 1997 \hfill ITP--UH--28/96 \\
\phantom{X}\hfill gr-qc/9612062

\vskip 1.0cm

\begin{center}

{\Large\bf Scalar Deformations of Schwarzschild Holes} 
\vskip 0.4cm
{\Large\bf and Their Stability}\\

\vskip 1.5cm

{\large Helge Dennhardt ${}^2$ \ and \ Olaf Lechtenfeld ${}^{1,2}$}

\vskip 1.0cm

{\it ${}^1$ School of Natural Sciences}\\
{\it Institute for Advanced Study}\\
{\it Olden Lane, Princeton, NJ 08540, U.S.A.}\\

\vskip 0.3cm
and
\vskip 0.3cm

{\it ${}^2$ Institut f\"ur Theoretische Physik}\\ 
{\it Universit\"at Hannover}\\
{\it Appelstra\ss{}e 2, 30167 Hannover, Germany}\\
{http://www.itp.uni-hannover.de/\~{}lechtenf/}\\

\vskip 2cm
\textwidth 6truein
{\bf Abstract}
\end{center}

\begin{quote}
\hspace{\parindent}
{}\ \ \
We construct two solutions of the minimally coupled Einstein--scalar field
equations, representing regular deformations of Schwarzschild black holes by a
self-interacting, static, scalar field. One solution features an exponentially
decaying scalar field and a triple-well interaction potential; the other one 
is completely analytic and sprouts Coulomb-like scalar hair. Both evade the 
no-hair theorem by having partially negative potential, in conflict with the
dominant energy condition. The linear perturbation theory around such 
backgrounds is developed in general, and yields stability criteria in
terms of effective potentials for an analog Schr\"odinger problem.
We can test for more than half of the perturbation modes, and our solutions
prove to be stable against those.
\end{quote}

\vfill

\textwidth 6.5truein
\hrule width 5.cm
\vskip.1in

{\small
\noindent ${}^2$
permanent address
}

\eject
\end{titlepage}
\newpage


\mysection{Introduction}
It has been known for a long time that black holes are notoriously
difficult to deform. This fact is captured by so-called no-hair theorems,
which classify stationary, asymptotically flat, regular, black-hole
solutions by a few conserved charges such as mass, angular momentum,
electric and magnetic charges.
Like any other no-go conjectures, no-hair theorems rest on certain
assumptions, and it is worthwile to investigate those for necessity.

The simplest case is that of ``scalar hair'' for Schwarzschild holes, 
where a gravitationally coupled static scalar field~$\Phi$ 
is subject to some self-interaction potential~$V(\Phi)$.  
For a recent review, see~\cite{Bekenstein1}.
The scalar no-hair theorem~\cite{Heusler,Bekenstein2} rests on the
{\it dominant energy condition\/} for the energy-momentum tensor~$T_{ij}$,
which stipulates that the energy current $j^i=T^i_j u^j$ of the matter field 
should never be space-like.
For a minimally coupled scalar field in a spacetime of signature $(-+++)$, 
this translates to $V(\Phi)\ge0$~\cite{Heusler}.
Therefore, scalar deformations of Schwarzschild holes, if at all possible,
need regions with negative interaction potential.\footnote{
The scale is set by the absence of a cosmological constant, 
$V(\Phi(\infty))=0$.}

The subject of this paper is to present two classes of static, isotropic, 
and regular solutions to the fully gravitating Einstein-scalar system, 
which circumvent the scalar no-hair theorem by virtue of having partially 
negative interaction potential. One of these solutions, with exponentially 
decaying scalar field, was given by one of the authors in an earlier paper
\cite{Bechmann}. The second solution, featuring a Coulomb-like scalar field,
is new and (in contrast to the first) completely analytical.

After restating the starting point equations in Section~2, we 
discuss both classes of solutions in some detail throughout Section~3.
In the second half of the paper, Section~4 provides a linear stability
analysis of Schwarzschild hole deformations by general scalar fields.
Following the techniques of Regge and Wheeler~\cite{Regge},
Vishveshwara~\cite{Vishveshwara}, and Zerilli~\cite{Zerilli},
we cast the perturbation equations into Schr\"odinger-like form
and read off stability criteria in terms of effective potentials.
Section~5 finally specializes and applies the results to our two
solutions, to decide upon their stability, before we conclude.


\mysection{The Equations}
The gravitational system we study in this paper is given by the
Einstein-Hilbert-Klein-Gordon action
\begin{equation}
\mS\ =\ \int d^4x\,\sqrt{-g}\,\bigl[ 
R + \frac12 g^{ij}\partial_i\Phi\partial_j\Phi+V(\Phi) \bigr]
\end{equation}
for a self-interacting scalar field minimally coupled to standard gravity.
The action is extremized by 
\begin{align}
0\ &=\ \D^i\partial_i\Phi-\partial_\Phi V(\Phi)\quad, \nonumber \\
0\ &=\ R_{ij}+\frac12\partial_i\Phi\partial_j\Phi+\frac12 g_{ij}V(\Phi)\quad.
\label{eomcov}
\end{align}
We specialize to spherical coordinates $x^i=(t,r,\vartheta,\varphi)$.
In the isotropic and static case all field degrees of freedom are functions
of the radial coordinate $r$ only, and the metric can be reduced to two
functions by residual coordinate transformations.
The field configuration is then given by~\footnote{
$R(r)$ is not to be confused with the Ricci scalar.}
\begin{equation}
\Phi\ =\ \Phi(r) \qquad\text{and}\qquad
ds^2\ =\ -G(r)\, dt^2 + G(r)^{-1} dr^2 + R(r)^2 d\Omega^2\quad.
\end{equation}
Instead of the usual Schwarzschild or isotropic coordinates,
we have chosen the different but very useful gauge $g_{tt}g_{rr}=-1$.
The equations (\ref{eomcov}) reduce to three independent ordinary
second-order differential equations,
\begin{align}
R''  \  &=\ \frac{1}{4}{\Phi'}^2R \quad,\nonumber \\
G''  \  &=\ \frac{1}{2}G{\Phi'}^2+2\frac{G{R'}^2-1}{R^2} \quad,\nonumber \\
V(\Phi)\ &=\ \frac{1}{2}G{\Phi'}^2+2\frac{R'G'}{R}+2\frac{G{R'}^2-1}{R^2}\quad,
\label{eom}
\end{align}
where the prime denotes a derivative with respect to~$r$.
The equation of motion for $\Phi$ follows from these.

Before we construct solutions of (\ref{eom}), it is necessary to put
boundary conditions for small and large values of~$r$.
Near spatial infinity we take the metric to approach the Schwarzschild
solution, i.e.
\begin{equation}\label{SS}
G\quad\xrightarrow{r\to\infty}\quad 1 - \frac{2M}{r} \qquad\text{and}\qquad
R\quad\xrightarrow{r\to\infty}\quad r \quad,
\end{equation}
which, using (\ref{eom}), implies that
\begin{equation}\label{decay}
\Phi\quad\xrightarrow{r\to\infty}\quad\Phi_0 + O(r^{-\frac12})
\qquad\text{with}\qquad V(\Phi_0) = 0 \quad.
\end{equation}
This precludes a cosmological constant.
Linear stability of the asymptotic Schwarzschild metric yields the condition
\begin{equation}
\partial^2_\Phi V(\Phi_0)\ \ge 0\ \quad.
\end{equation}
For small $r$, we like to encounter a black hole, signified by the presence
of an event horizon at some value $r=h$ where
\begin{equation}
G(h)\ =\ 0 \qquad\text{and}\qquad  G(r>h)\ >\ 0 \quad.
\end{equation}
The black-hole singularity at $r=s$ is given by the (right-most) pole of~$G$,
\begin{equation}
G(r)\quad\xrightarrow{r\to s}\quad\pm\infty \quad.
\end{equation}

It proves useful to formally integrate the second equation of (\ref{eom}).
With (\ref{SS}), one arrives at
\begin{equation}\label{formalG}
G(r)\ =\ R(r)^2\int_r^\infty\! d\Tilde{r}\;\frac{2\Tilde{r}-6M}{R(\Tilde{r})^4}
\quad.
\end{equation}
The integration constant $M$ is chosen to agree with the black-hole mass
in (\ref{SS}).\footnote{
If $\Phi$ decays as $r^{-1}$, then $R\sim r+\text{constant}$,
and the two definitions of $M$ differ by a constant.}
Given a partial solution $(\Phi,R)$ of (\ref{eom}), it remains to evaluate
the integral above.


\mysection{Two Solutions}
In this section we will present two families of solutions to (\ref{eom}),
one with Coulombic and one with exponential scalar field. The Coulombic
solution is completely analytic and, to our knowledge, new to the literature.
The exponential solution is partly analytic and was given earlier by one
of the authors~\cite{Bechmann}.
If the potential is set to zero, the equations (\ref{eom}) are soluble
and yield (for fixed horizon) a known one-parameter family of solutions
\cite{Buchdahl,Bechmann}, with $\Phi(r)\sim\ln(1-h/r)$. Only for one value
of the parameter one avoids the logarithmic singularity, at the price of
setting $\Phi=0$ and recovering the Schwarzschild metric.
For a general potential $V(\Phi)$, the equations (\ref{eom}) cannot be solved
analytically. We therefore approach the problem in reverse:
\begin{itemize}
\item 
We make an ansatz for $\Phi$, choosing a class of functions
\item
We determine $R$ from the first of (\ref{eom}). One integration constant
is gauged away, the second one gets fixed by asymptotic flatness.
\item
Inserting $R$ into (\ref{formalG}), we obtain $G$. The mass $M$ appears
as an additional parameter.
The right-most zero of $G$ gives the event horion.
\item
The third of (\ref{eom}) yields $U(r)=V(\Phi(r))$. To find $V(\Phi)$, 
we invert $\Phi(r)$ to $r(\Phi)$.
\item
The functions $R(r)$, $G(r)$, and $V(\Phi)$ are plotted.
\end{itemize}


\subsection{Coulombic Scalar Field}
We assume that the scalar field has power-like behavior,
\begin{equation}
\Phi(r)\ =\ q\cdot r^\alpha \quad,
\end{equation}
with the restriction $\alpha\le-\frac12$, due to (\ref{decay}).
The resulting equation for $R$ (\ref{eom}),
\begin{equation}
R''\ =\ \frac{1}{4}q^2\alpha^2 r^{2\alpha-2}R \quad,
\end{equation}
can be solved in terms of Bessel functions,
\begin{align}\label{Bessel}
R(r)\ &=\ \sqrt{r}Z_\frac{1}{2\alpha}
\left(i\frac{\pm q}{2}\cdot r^\alpha\right) \nonumber \\
&=\ \sqrt{r}\left[
c_1\cdot I_\frac{1}{2\alpha}\left(\frac{\pm q}{2}\cdot r^\alpha\right)+
c_2\cdot K_\frac{1}{2\alpha}\left(\frac{\pm q}{2}\cdot r^\alpha\right)
\right]\quad.
\end{align}
The choice of sign originates in the $\Phi\to-\Phi$ symmetry of the equations
and will be hidden in the new coupling $Q:=\pm q$.

In order to be able to proceed analytically, 
it is necessary to specialize further.
Bessel functions become elementary for $\frac{1}{2\alpha}=n+\frac12$, with
$n\in\Z$, but the above restriction leaves only $\alpha=-1$,
i.e. Coulombic fall-off for $\Phi$. Hence, we confine ourselves to the case
\begin{equation}
\Phi\ =\ \frac{q}{r} \quad. \label{PPolynom}
\end{equation}
Then, the solution (\ref{Bessel}) turns into
\begin{equation}\label{Rcoulgen}
R(r) \ =\  c_3\cdot r\left(e^\frac{Q}{2r}+e^{-\frac{Q}{2r}}\right)
       +c_4\cdot r e^{-\frac{Q}{2r}} \quad,
\end{equation}
with new integration constants $c_3$ and $c_4$.
We rescale time and radial coordinates by
\begin{equation}
r\ =\ \bar{r}/ |c_4| \qquad\text{and}\qquad t\ =\ \bar{t}\cdot |c_4| 
\end{equation}
and absorb its effect on $\Phi,G,R,$ and $V$ by redefining
\begin{equation}
q\ =\ \bar{q}/|c_4| \quad,\quad
M\ =\ \bar{M}/|c_4| \quad,\quad
c_3\ =\ \bar{c}_3\cdot|c_4|
\end{equation}
so that $\bar{c}_4=\pm1$.
Dropping the bars, $R$ of (\ref{Rcoulgen}) (with $c_4=\pm1$) tends to
\begin{equation}
R \quad\xrightarrow{r\to\infty}\quad  c_3\cdot 2r\pm r 
  \quad\xrightarrow{!}\quad r\quad,
\end{equation}
which allows only $c_3=0$ or $c_3=1$, depending on the sign.
In both cases, the result is~\footnote{
We may absorb a sign in the exponent into $Q$, since $Q$ is
defined only up to a sign.}
\begin{equation}\label{Rcoul}
R(r)\ =\ r\cdot e^{-\frac{Q}{2r}} \quad.
\end{equation}

\vspace{0.4cm}
\figure[ht!]
\begin{center}\
\epsfbox[0 0 200 200]{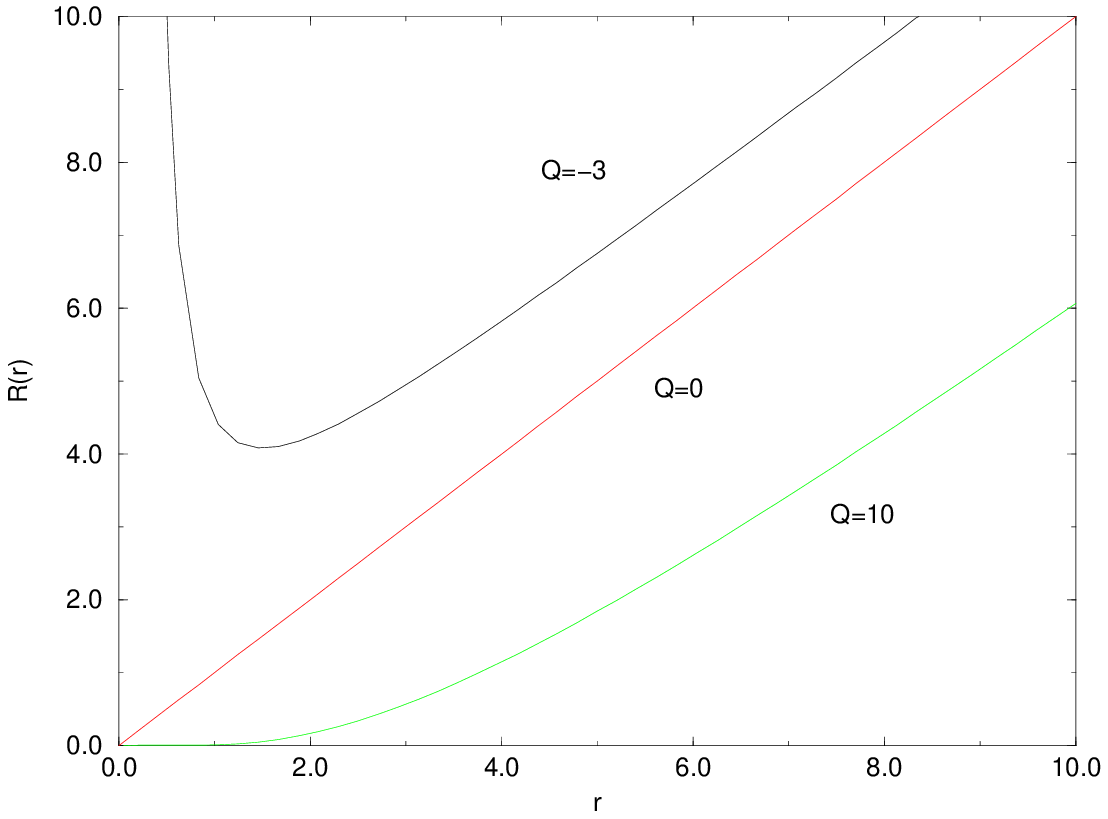}
\caption{The function $R(r)$ for three values of $Q$}
\label{figure1}
\end{center}
\endfigure
\vspace{-0.2cm}

We must distinguish three cases:
\begin{itemize}
\item 
$Q>0$: $R$ is a good radial coordinate, since $R(r)$ is positive and monotonic.
\item 
$Q=0$: The Schwarzschild case ($\Phi=0$).
\item 
$Q<0$: $R>0$ everywhere, but $R(r)$ is not monotonic.
\end{itemize}

Inserting (\ref{Rcoul}) into (\ref{formalG}) and evaluating the integral,
one arrives at
\begin{equation}
G(r)\ =\ \Bigl(
-\frac{Q+3M}{2Q^3}\cdot r^2+\frac{Q+3M}{Q^2}\cdot r -\frac{3M}{Q}
\Bigr)\cdot e^\frac{Q}{r}\ +\ \frac{Q+3M}{2Q^3}r^2\cdot e^{-\frac{Q}{r}}\quad.
\end{equation}
The asymptotic behavior,
\begin{equation}
G\quad\xrightarrow{r\to\infty}\quad 
1-\frac{2M-\frac{1}{3}Q}{r}+O\left(\frac{1}{r^2}\right) \quad,
\end{equation}
signifies that the black-hole mass is actually
\begin{equation}
\Tilde{M}\ =\ M-\frac{1}{6}Q \quad,
\end{equation}
in terms of which the metric function reads
\begin{equation}
G(r)\ =\ \Bigl(
-\frac{3Q+6\Tilde{M}}{4Q^3}\cdot r^2+\frac{3Q+6\Tilde{M}}{2Q^2}\cdot r 
-\frac{Q+6\Tilde{M}}{2Q}\Bigr)\cdot e^\frac{Q}{r}\
+\ \frac{3Q+6\Tilde{M}}{4Q^3}r^2\cdot e^{-\frac{Q}{r}} \quad.
\end{equation}

\vspace{0.5cm}
\figure[ht!]
\begin{center}\
\epsfbox[0 0 200 200]{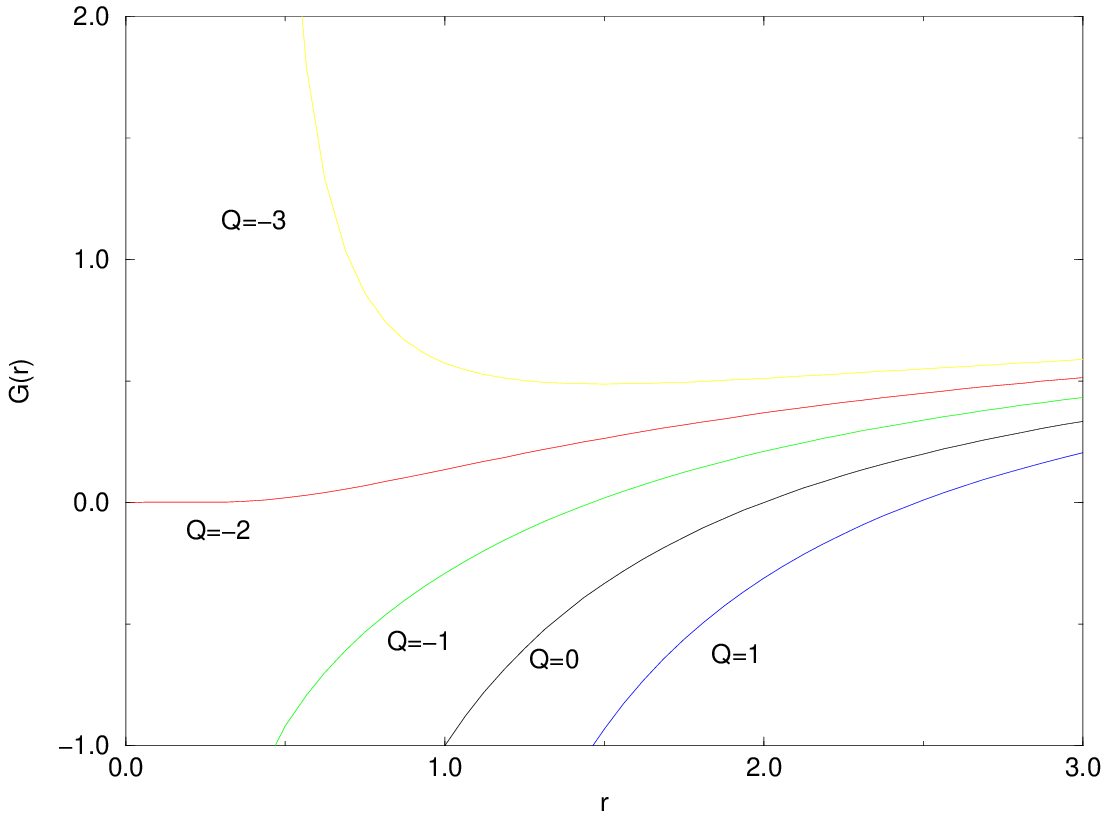}
\caption{The function $G(r)$ for $\protect\Tilde{M}=1$ and three values of $Q$}
\label{figure2}
\end{center}
\endfigure

Again, a case distinction should be made:
\begin{itemize}
\item
$Q>0$: the scalar field increases the size of the black hole.
\item
$Q=0$: the Schwarzschild solution.
\item
$-2\Tilde{M}<Q<0$: the scalar field decreases the size of the black hole.
\item
$Q\le-2\Tilde{M}$: a naked singularity, since $G\xrightarrow{r\to0}+\infty$
	without a horizon.
\end{itemize}
Therefore, we restrict ourselves to 
\begin{equation}
Q\ >\ -2\Tilde{M} \quad.
\end{equation}

Finally, we employ the third of (\ref{eom}) to compute the potential,
\begin{equation}
U(r)\ =\ -\frac{3(Q+2\Tilde{M})}{Q^3}\cdot \left[
\left(3+\frac{Q^2}{r^2}\right)\cdot\sinh\left(\frac{Q}{r}\right)
-3\frac{Q}{r}\cdot\cosh\left(\frac{Q}{r}\right)\right] \quad,
\end{equation}
and invert via $r=q/\Phi$ to obtain
\begin{equation}\label{Vcoul}
V(\Phi)\ =\ -\lambda \cdot \Bigl[
\left(3+\Phi^2\right)\cdot\sinh|\Phi|-3|\Phi|\cdot\cosh\Phi\Bigr] \quad,
\end{equation}
with
\begin{equation}
\lambda\ :=\ \frac{3(Q+2\Tilde{M})}{|Q|^3}\qquad > 0 \quad.
\end{equation}
Since $\Phi=q/r$ reaches only positive (or only negative) values of~$\Phi$,
(\ref{Vcoul}) is valid only for one sign of $\Phi$.
It is, however, natural to impose a $\Phi\to-\Phi$ symmetry, 
and (\ref{Vcoul}) respects it already.

\vspace{0.7cm}
\figure[ht!]
\begin{center}\
\epsfbox[0 0 200 200]{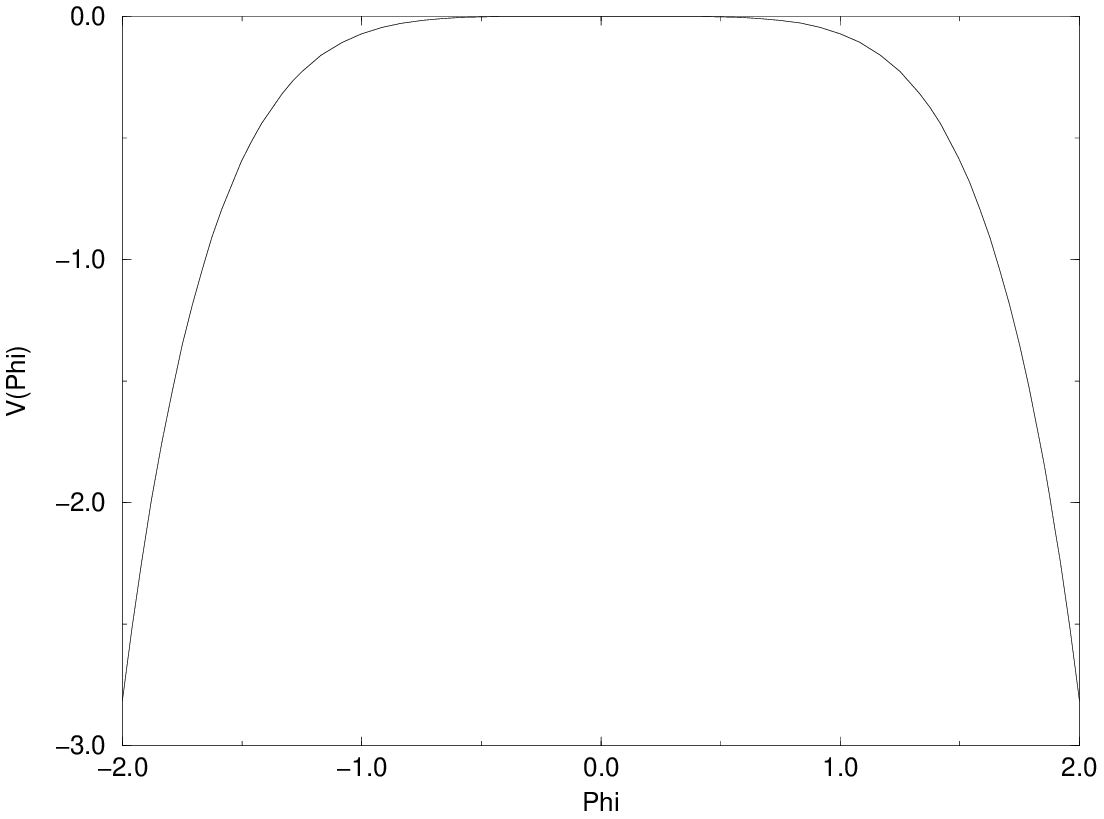}
\caption{The function $V(\Phi)$ for $\lambda=1$}
\label{figure3}
\end{center}
\endfigure
\vspace{-0.2cm}

The potential function $V(\Phi)$ is not analytic at the origin,
but varies as
\begin{equation}
V(\Phi)\quad\xrightarrow{\Phi\to0}\quad-\frac{\lambda}{15}\cdot|\Phi|^5
\end{equation}
so that $\partial_\Phi^2V(0)=0$ and the vacuum stability is marginal.
It is apparent that the interaction potential is everywhere negative.
Therefore, the no-hair theorem \cite{Heusler,Bekenstein2} simply does not 
apply to our solution since the latter does not obey the dominant energy 
condition.


\subsection{Exponential Scalar Field}
We assume that the scalar field drops exponentially,
\begin{equation}
\Phi(r)\ =\ q\cdot e^{-mr} \quad,
\end{equation}
with $m\ge0$ for asymptotic flatness.
This ansatz and the solution presented in the following
were the subject of \cite{Bechmann}.

Again, the resulting equation for $R$,
\begin{equation}
R''(r)\ =\ \frac{1}{4}q^2m^2e^{-2mr}R(r) \quad,
\end{equation}
can be solved in terms of Bessel functions, namely
\begin{align}
R(r)\ &=\ Z_0 \left( i\frac{\pm q}{2}\cdot e^{-mr} \right) \nonumber \\
&=\ c_1\cdot I_0 \left( \frac{\pm q}{2}\cdot e^{-mr} \right)
 + c_2\cdot K_0 \left( \frac{\pm q}{2}\cdot e^{-mr} \right) \quad,
\label{Rexpgen}
\end{align}
with two integration constants $c_1$ and $c_2$.
Here too, we rescale time and radial coordinates by
\begin{equation}
r\ =\ \bar{r}/m \qquad\text{and}\qquad t\ =\ \bar{t}\cdot m
\end{equation}
and absorb its effect on $\Phi,G,R,$ and $V$ by redefining
\begin{equation}
M\ =\ \bar{M}/m
\end{equation}
so that $\bar{m}=1$.
With a second coordinate transformation ($\gamma$ is Euler's constant),
\begin{equation}
\bar{r}\ =\ \hat{r} + \ln\left|\frac{q}{4}\right|
+ \gamma - \frac{c_1}{c_2} \quad,
\end{equation}
we get rid of~$c_1$,
but must redefine
\begin{align}
\bar{M}\ &=\ \hat{M}+\frac{1}{3}\cdot\left(\ln\left|\frac{q}{4}\right| 
    +\gamma-\frac{c_1}{c_2}\right) \quad, \nonumber \\
\hat{q}\ &=\ \sgn(q)\cdot 4\,e^{c_1/c_2-\gamma} \quad.
\end{align}
Dropping the hats, $R$ of (\ref{Rexpgen}) (with $m=1$ and $c_1=0$)
tends to
\begin{equation}
R \quad\xrightarrow{r\to\infty}\quad c_2\cdot r
  \quad\xrightarrow{!}\quad r\quad,
\end{equation}
so that we have to take $c_2=1$.
Inserting the infinite series representation of $K_0$,
the radial function may be expressed as
\begin{equation}\label{Rexp}
R(r)\ =\ r+\sum_{k=1}^\infty \Bigl(r+\sum_{\ell=0}^{k}\frac{1}{\ell}\Bigr)
         \cdot\frac{1}{(k!)^2}\left(\frac{q}{4}e^{-r}\right)^{2k} \quad.
\end{equation}
The sign ambiguity of (\ref{Rexpgen}) has disappeared since only $q^2$ enters.
\vfill\eject

\figure[ht!]
\begin{center}\
\epsfbox[0 0 200 200]{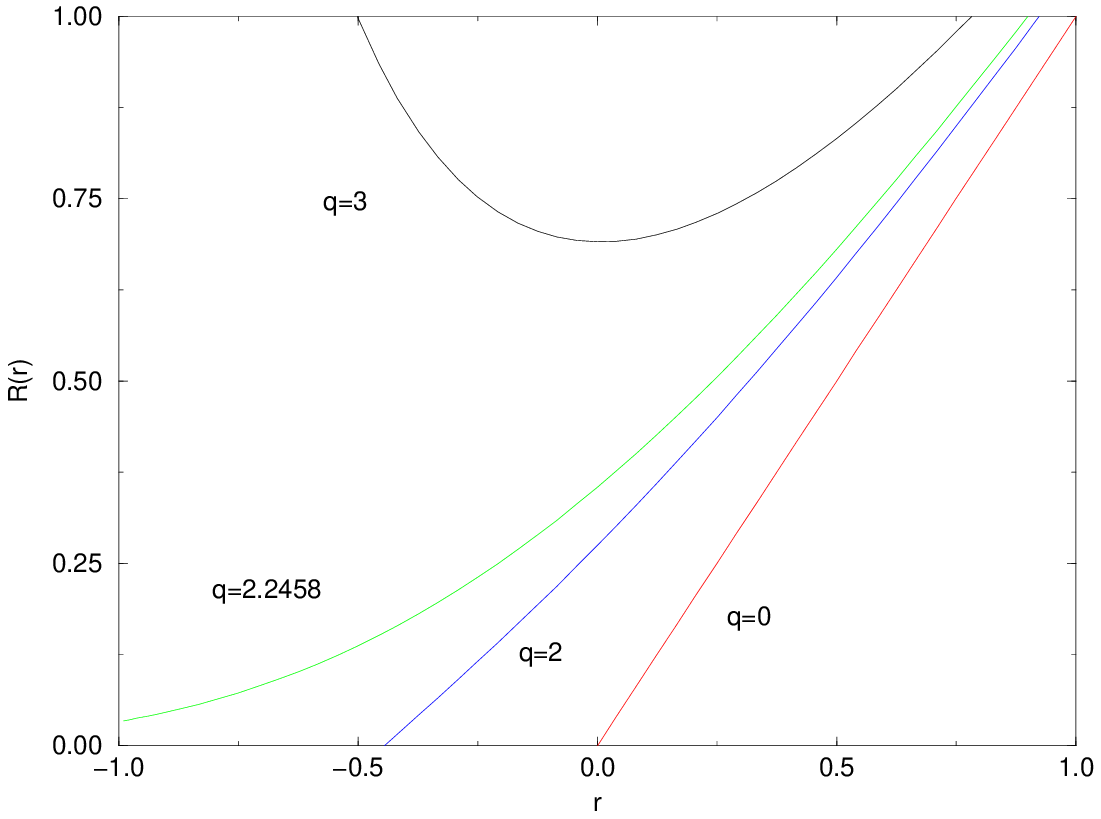}
\caption{The function $R(r)$ for four values of $q$}
\label{figure4}
\end{center}
\endfigure

Three cases are to be distinguished:
\begin{itemize}
\item
$|q|>4\exp(-\gamma)$: $R>0$ everywhere, but $R(r)$ is not monotonic.
\item
$0<|q|\le4\exp(-\gamma)$: $R$ is a good radial coordinate, 
	since $R(r)$ is positive and monotonic.
\item
$|q|=0$: The Schwarzschild case ($\Phi=0$).
\end{itemize}

Inserting (\ref{Rexp}) into (\ref{formalG}), one discovers that the integral
cannot be done analytically. It is sensible, however, to expand the
integrand in powers of $q/4\cdot\exp(-r)$ and integrate it term by term.
The result is the series
\begin{equation}\label{Gexp}
G(r)\ =\ \sum_{k=0}^\infty\Bigl[a_k(r,M)+\sum_{\ell=1}^\infty b_{k\ell}(r,M)
     \cdot q^{2l}\Ei(2\ell r)\Bigr]\cdot q^{2k}\,e^{-2kr} \quad,
\end{equation}
where the coefficients $a_k$ and $b_{k\ell}$ are linear in $M$ and
polynomial in $r$ and $\frac1r$.
$\Ei(x)=\int_x^\infty\!dy\,e^{-y}/y$ denotes the exponential-integral function.
Asymptotically, (\ref{Gexp}) yields the Schwarzschild metric with mass~$M$.
The following graphs were obtained by keeping terms up to $k=\ell=6$,
i.e. $O(\exp(-12r))$.

\figure[ht!]
\begin{center}\
\epsfbox[0 0 200 200]{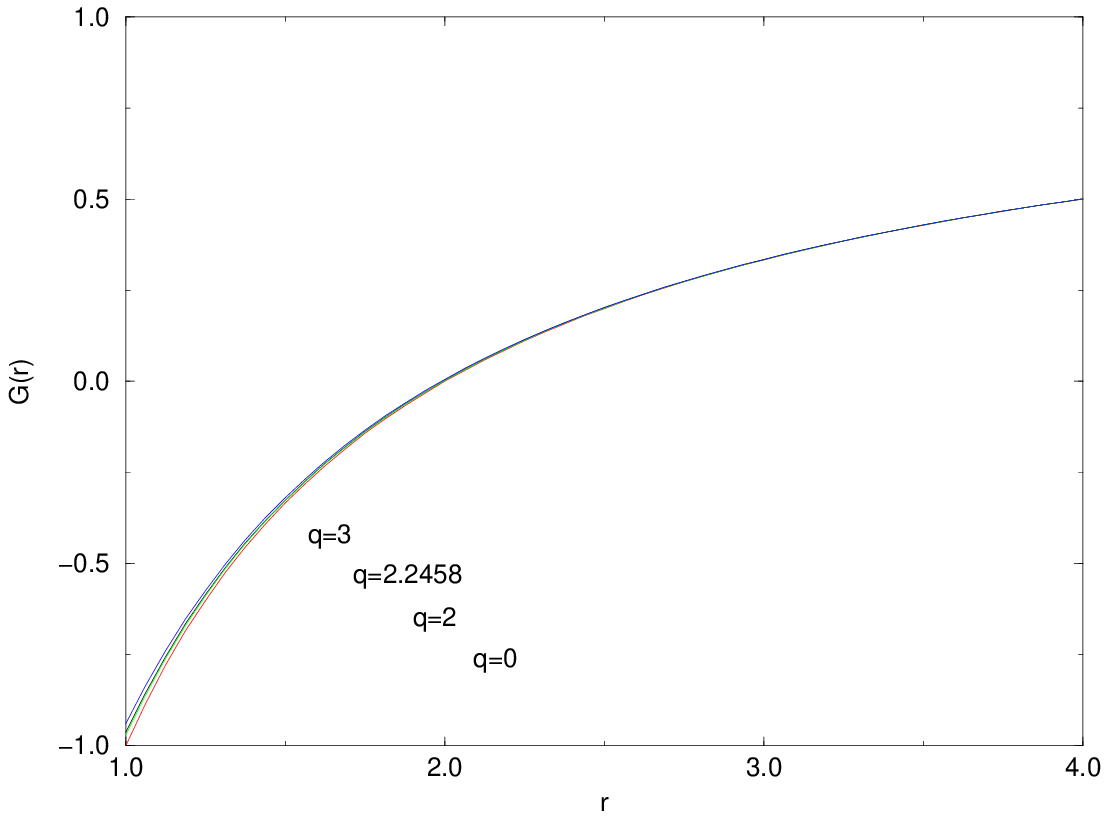}
\caption{The function $G(r)$ for $M=1$ and four values of $q$}
\label{figure5}
\end{center}
\endfigure

\figure[ht!]
\begin{center}\
\epsfbox[0 0 200 200]{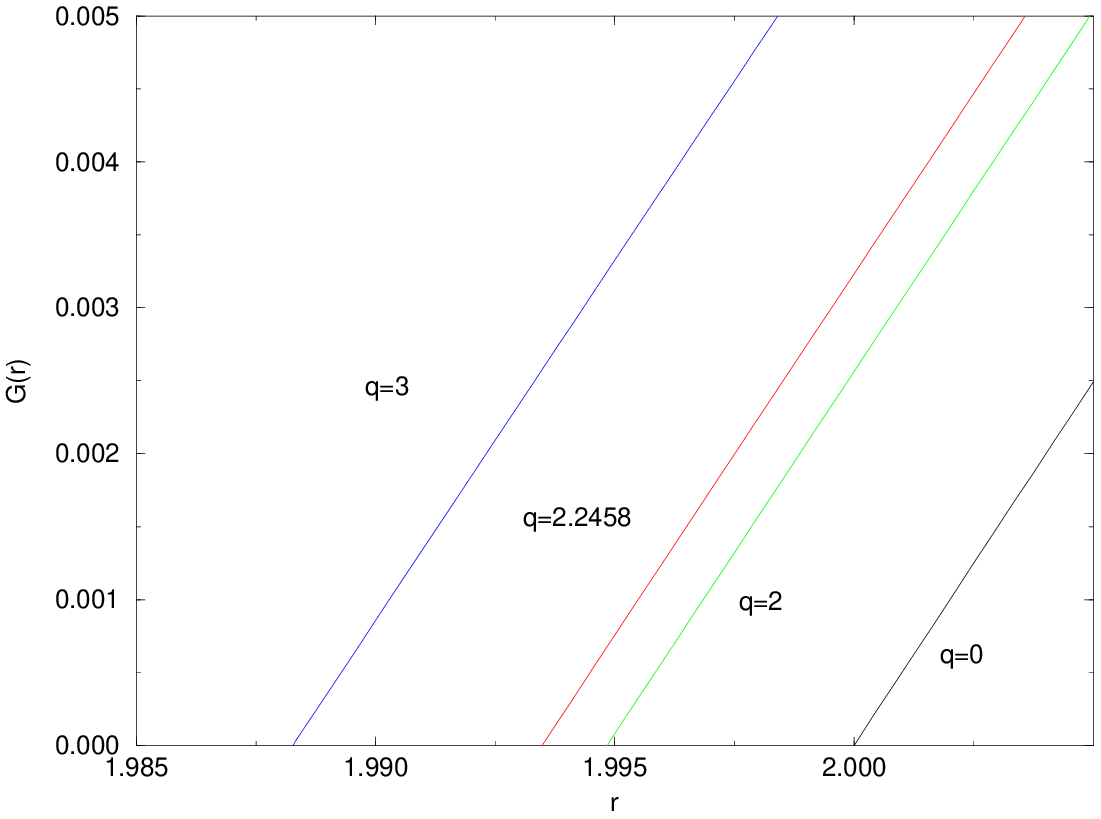}
\caption{Detail of Figure~\ref{figure5} near the horizon}
\label{figure6}
\end{center}
\endfigure

Despite the fact that $R$ depends greatly on $q$
(see Figure~\ref{figure4}), the metric function $G$ hardly varies with $q$.
Magnifying Figure~\ref{figure5} brings out a small shift of the horizon.
Note that, contrary to the Coulombic case, the singularity is shielded
for all values of~$q$.

We finally come to the potential. 
Carrying over the series expansion to the third of (\ref{eom}),
one gets an expression of the same form as (\ref{Gexp}),
\begin{equation}
U(r)\ =\ \sum_{k=1}^\infty\Bigl[ c_k(r,M)+\sum_{\ell=1}^\infty d_{k\ell}(r,M)
     \cdot q^{2\ell}\Ei(2\ell r) \Bigr] \cdot q^{2k}\,e^{-2kr} \quad,
\end{equation}
where $c_k$ and $d_{k\ell}$ share the properties of $a_k$ and $b_{k\ell}$.
Inverting by $r=-\ln|\Phi/q|$ we find
\begin{equation}\label{Vexp}
V(\Phi)\ =\ \sum_{k=1}^\infty \Bigl[ c'_k(|\Phi|,M)
+\sum_{\ell=1}^\infty d'_{k\ell}(|\Phi|,M)\cdot q^{2\ell}\Ei(-2\ell\ln|\Phi/q|)
\Bigr] \cdot \Phi^{2k} \quad,
\end{equation}
where $c'_k(|\Phi(r)|,M)=c_k(r,M)$ 
and $d'_{k\ell}(|\Phi(r)|,M)=d_{k\ell}(r,M)$.
Again, it is natural to extend the solution for only positive 
(or only negative) values of~$\Phi$ to the whole real axis by
taking $V(-\Phi)=V(\Phi)$, already made explicit in (\ref{Vexp}).

\figure[ht!]
\begin{center}\
\epsfbox[0 0 200 200]{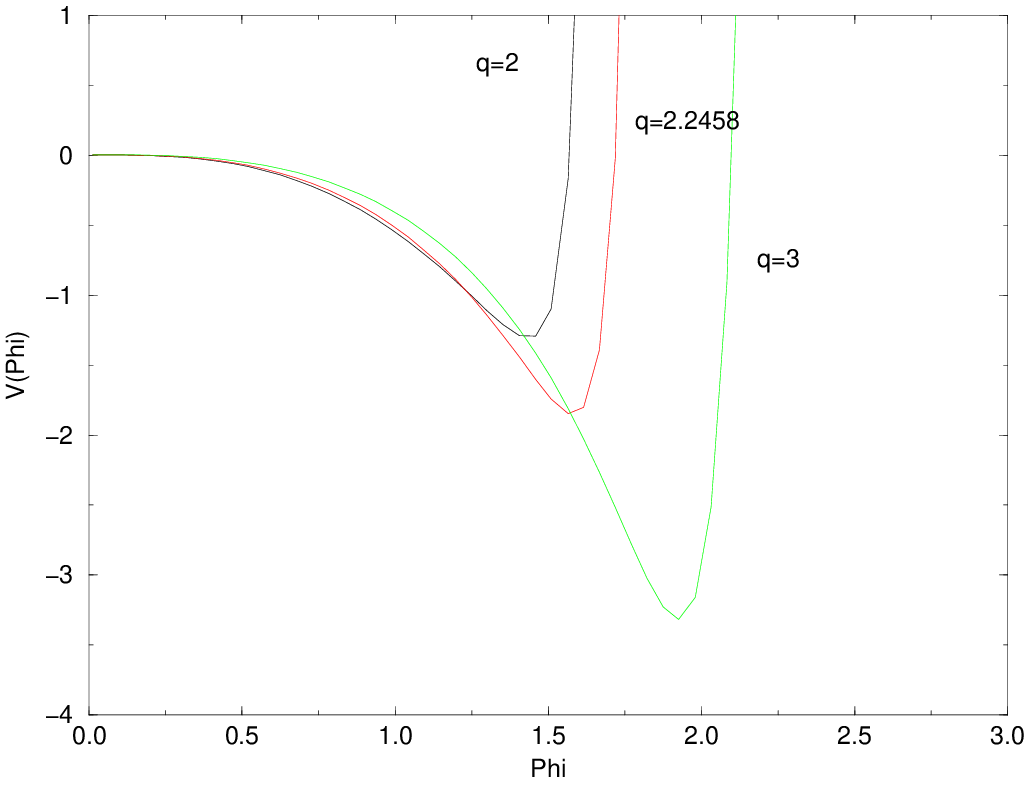}
\caption{$V(\Phi)$ for $M=1$ and three values of $q$}
\label{figure7}
\end{center}
\endfigure

\figure[ht!]
\begin{center}\
\epsfbox[0 0 200 200]{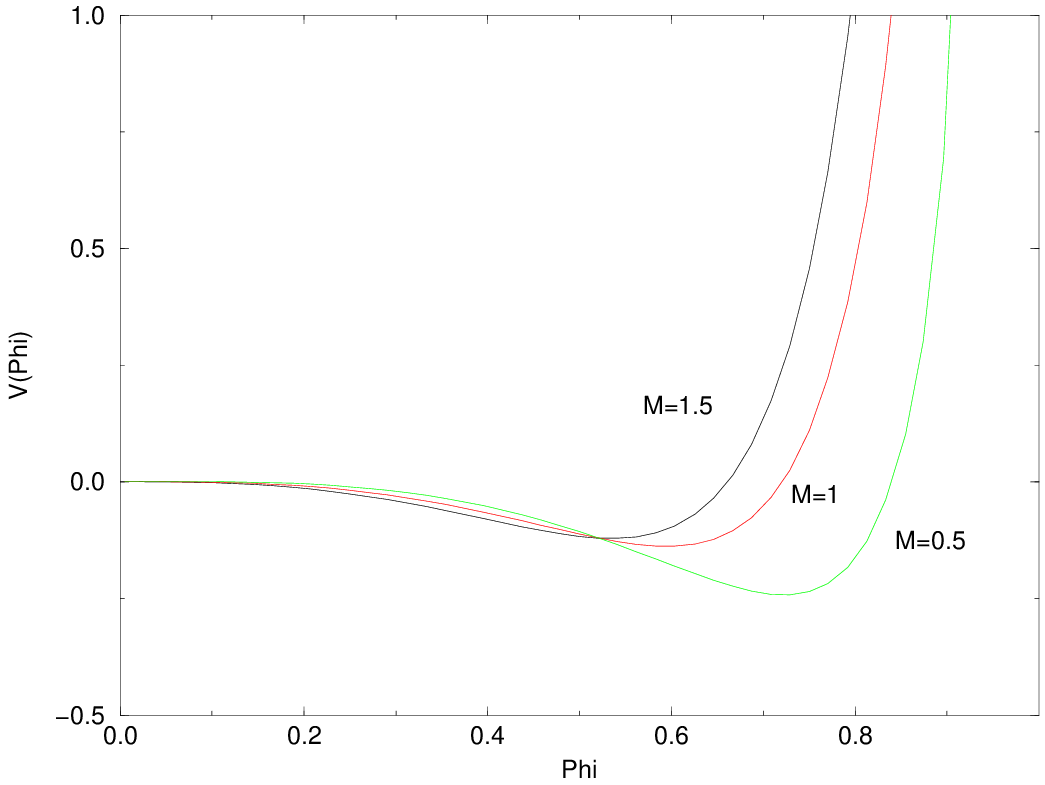}
\caption{$V(\Phi)$ for $q=1$ and three values of $M$}
\label{figure8}
\end{center}
\endfigure

Taking a closer look at the origin reveals the local minimum
which also follows from
\begin{equation}
V(\Phi) \quad\xrightarrow{\Phi\to 0}\quad c_1\cdot \Phi^2\ =\
    \frac{1}{2}\Phi^2-\frac{2M\Phi^2}{\ln|\Phi/q|} 
\end{equation}
and, again, is nonanalytic but smooth (even $\C^\infty$). 
So, contrary to the Coulombic solution, vacuum stability is manifest.

\figure[ht!]
\begin{center}\
\epsfbox[0 0 200 200]{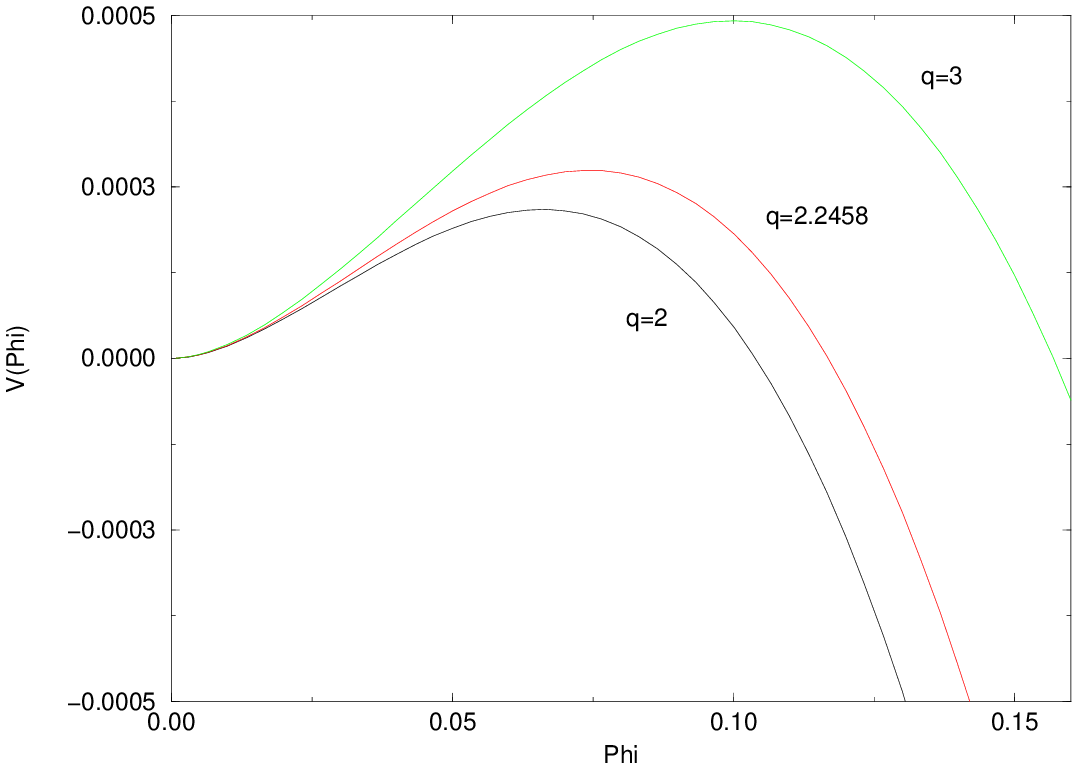}
\caption{Detail of Figure \ref{figure7} near the origin}
\label{figure9}
\end{center}
\endfigure

The no-hair theorem \cite{Heusler,Bekenstein2} gets circumvented 
by having regions of negative potential, 
again violating the dominant energy condition.
In \cite{Bechmann}, the energy-momentum tensor of this configuration
was displayed. Near the horizon, the energy density 
$\rho=V+\frac12 G{\Phi'}^2$ drops below zero, which violates the weak
energy condition as well.


\mysection{Perturbations}

To examine the stability of the static solutions introduced in Section~3, 
we first decompose possible perturbations into
tensorial modes and simplify them with a gauge transformation and
a rotation. We then use this general ansatz for the perturbations 
to linearize the equations of motion around the static solution and
put them into a Schr\"odinger-like form. In this form, instability is 
equivalent to the existence of negative eigenvalues of the Schr\"odinger
problem, a fact which leads to a general stability condition for static 
solutions.

This procedure for dealing with stability questions in general relativity 
was first described in 1957 by Regge and Wheeler~\cite{Regge}, who
proved the stability of black holes against odd parity perturbations.
The proof for even-parity perturbations was achieved in 1969 by
Vishveshwara \cite{Vishveshwara}, using Kruskal coordinates. The work of
Zerilli \cite{Zerilli} completed these examinations.

In this paper, we are dealing with a more general system containing an 
additional scalar field, which implies that it is always possible to get the 
equations of Regge, Wheeler, Vishveshwara, and Zerilli by carrying out the 
zero scalar field limit ($\Phi\to0, V(\Phi)\to0,G\to 1-\frac{2M}{r},R\to r$).


\subsection{Canonical Ansatz}

To get a general ansatz for the perturbations of our system, we first
separate the perturbations from the background
\begin{equation} 
  \begin{split}
    g_{ij}^{tot}\ & =\ g_{ij}+ \delta g_{ij}\\
    \Phi^{tot}\   & =\ \Phi + \delta\Phi
  \end{split}
  \label{a1}
\end{equation}
with $\delta g_{ij}$ and $\delta\Phi$ being infinitesimal small.

Described in this way, the perturbations are independent fields, living 
on a static, spherically symmetric background, which implies constancy of
energy, angular momentum, and parity. Because of this, modes of different
energy, angular momentum, or parity do not mix when time evolves, and
each mode can be examined separately for stability.

The decomposition into modes with fixed energy is achieved by introducing 
a factor $\exp(i\omega t)$ in each mode. Likewise, the decomposition into 
modes with fixed angular momentum yields factors $Y_L^M(\vartheta,\varphi)$.
This leaves a pure radial part for the remaining degrees of freedom, which
then is split into two parts with opposite parity.

The functions $Y_L^M$ already have definite parity $(-1)^L$, so the
parity separation is not necessary for the scalar perturbations $\delta\Phi$. 
In contrast, the metric perturbations $\delta g_{ij}$ with their
tensorial character bear directional information which does not remain 
unchanged under a parity transformation and hence must be split carefully
into two parts. This is shown i.e. in \cite{Regge}.

The resulting formulas can then be simplified enormously by a gauge fixing,
which eliminates four radial degrees of freedom. Furthermore, a rotation 
is performed for each mode, which sets $M$ to zero and thus transforms the 
functions $Y_L^M$ into Legendre polynomials $P_L(\cos\vartheta)$
so that all dependence on $\varphi$ disappears. 
Regge and Wheeler  called the resulting 
perturbation modes canonical\footnote{To get the notation of \cite{Regge}, 
replace $h_1\to H_0\cdot G$, $h_2\to H_0$, $h_3\to H_2/G$, 
$h_7\to K\cdot R^2$, $h_8\to h_0$, $h_9\to h_1$}. They read

for even parity:
\vspace{-0.3cm}
\begin{align}
  \delta g_{ij}\ & =\ \begin{pmatrix}
               h_1(r)         & h_2(r)         & 0   & 0 \\ 
               h_2(r)         & h_3(r)         & 0   & 0 \\
               0              & 0              & h_7(r) & 0 \\
               0              & 0              & 0   & h_7(r)\sin^2\vartheta
             \end{pmatrix} P_L(\cos\vartheta) e^{i\omega t} \quad,
             \nonumber\\\nonumber\\
  \delta\Phi\   & =\ \phi(r)\cdot  P_L(\cos\vartheta) e^{i\omega t} \quad,
  \label{a2}
  \\\nonumber
\end{align}  
for odd parity:
\vspace{-0.3cm}
\begin{align}
  \delta g_{ij}\ &=\ 
           \begin{pmatrix}
               0    & 0   & 0 & h_8(r)\\
               0    & 0   & 0 & h_9(r)\\
               0    & 0   & 0 & 0\\
                h_8(r) & h_9(r) & 0 & 0
             \end{pmatrix} 
             \cdot \sin\vartheta \;\partial_\vartheta P_L(\cos\vartheta) 
		e^{i\omega t} \quad,
             \nonumber\\\nonumber\\
  \delta\Phi\  &=\ 0 \quad,
  \label{a3}
\end{align}
where ``even'' and ``odd'' denote parity $(-1)^L$ and $(-1)^{L+1}$, 
respectively.


\subsection{Odd-Parity Sector}

To examine the odd-parity perturbations, we first insert
(\ref{a1}) with the canonical odd-parity ansatz (\ref{a3})
into the equations of motion (\ref{eomcov}) and retain only
terms linear in the perturbations $h_8$ and $h_9$.

Along this procedure, one encounters second and third order
derivatives of the Legrende polynomials, which can be reduced in
order by their defining property
\begin{equation}
  \partial_\vartheta^2 P_L(\cos\vartheta)\ 
    =\ -\frac{\cos\vartheta}{\sin\vartheta}\cdot\partial_\vartheta 
	P_L(\cos\vartheta) -L(L+1)\cdot P_L(\cos\vartheta)
\end{equation}
so that only derivatives of order 0 and 1 are left.
After tedious calculations one obtains two independent
differential equations,
\begin{align}
  0\ & =\ \left[\Bigl(\frac{\frac{1}{2}L(L+1)-1}{R^2}
	-\frac{1}{2}\frac{\omega^2}{G}\Bigr)\cdot h_9
        +i\omega\frac{R'}{GR}\cdot h_8
        -\frac{1}{2}\frac{i\omega}{G}\cdot h_8'\right]\cdot\nonumber\\
    &\quad\cdot\sin\vartheta\,\partial_\vartheta P_L(\cos\vartheta) 
	e^{i\omega t} \quad,
        \label{a4}\\
    &   \nonumber\\
  0\ & =\ \left[\frac{i\omega}{G}\cdot h_8-G'\cdot h_9
	-G\cdot h_9'\right]\cdot\nonumber\\
    & \quad \cdot \left(\frac{1}{2}L(L+1)\sin\vartheta\, P_L(\cos\vartheta)
        +\cos\vartheta\, \partial_\vartheta P_L(\cos\vartheta)\right)
        \cdot e^{i\omega t} \quad.
        \label{a5}
\end{align}
These are the elements (1,3) and (2,3) of the perturbation equations for
the gravitational field. Equation (\ref{a4}) was simplified by using
the static equations of motion (\ref{eom}) to eliminate $R''$ and $V$.

The element (0,3) of the graviational equations results in
a dependent differential equation, all other elements are given by symmetry 
or yield the identity. The equation for the scalar field also yields nothing, 
since the scalar perturbations do not take part in the odd-parity 
oscillations at all.
The system is hence fully described by the two differential
equations (\ref{a4}) and (\ref{a5}).

At this point, it is necessary to separate out two degenerate cases for low
values of the angular momentum $L$. They are
\begin{itemize}
  \item $L=0$: In this case, $P_0(\cos\vartheta)=1$, both differential
        equations (\ref{a4}) and (\ref{a5}) degenerate to the
        identity, and no odd-parity perturbation really exists, since 
        they vanish totally in our ansatz (\ref{a3}).
  \item $L=1$: In this case, $P_1(\cos\vartheta)=\cos\vartheta$ and equation
        (\ref{a5}) degenerates. An additional gauge transformation, 
        \begin{equation}
          x'{}^i\ =\ x^i+\delta^i{}_\varphi\frac{i}{\omega R^2} h_8 \cdot 
	  e^{i\omega t} \quad,
        \end{equation}
        and a redefinition of $h_9$ make $h_8$ vanish totally
        from our perturbation ansatz. The remaining differential equation
        (\ref{a4}) then states $h_9$ to be zero as well. Hence, the 
        perturbation has been gauged away completely.
\end{itemize}

This discussion shows that the lowest angular momentum of odd-parity 
perturbations is $L=2$, hence the angular parts of the differential equations 
never vanish. Since these equations have to hold for all angles and all times,
the radial parts (in square brackets) can be set to zero separately, leaving
a pure radial differential system.

Solving the radial part of (\ref{a5}) for $h_8$, one gets
\begin{equation}
   h_8\ =\ \frac{G}{i\omega}\cdot\left(G'\cdot h_9 + G h_9'
	\right) \quad.
  \label{a6}
\end{equation}
Inserting this into (\ref{a4}) yields
\begin{align}
\left(2\frac{GG'R'}{R}-{G'}^2-GG''-\frac{G}{R^2}\left(L(L+1)-2\right)\right)
\cdot h_9& \nonumber \\
+\left(2\frac{G^2R'}{R}-3GG'\right)\cdot h_9'-G^2\cdot h_9''&\
=\ \omega^2\cdot h_9 \quad.
\label{a7}
\end{align}
As one can see, equation (\ref{a7}) determines the entire dynamics,
while (\ref{a6}) only expresses $h_8$ in terms of the dynamical
variable $h_9$. 
For vanishing interaction potential, it reduces to the Regge-Wheeler 
equation~\cite{Regge}.  The remaining problem now consists of solving the 
second-order differential equation (\ref{a7}), which in fact is not
possible without specifying the static solution functions $G$ and $R$.

For stability questions, however, it is unnecessary to solve this differential
equation explicitely. We only need information about the spectrum of 
frequencies $\omega$. Because of the time evolution $\exp(i\omega t)$ of the 
perturbation modes, it is sufficient to determine whether $\omega$ has an 
imaginary part or not. To decide this question, we will put our
differential equation (\ref{a7}) into a Schr\"odinger-like form.
To this end, we introduce a new dynamical variable $\Psi$ via
\begin{equation}
  \Psi(r)\ :=\ \frac{G(r)}{R(r)}\cdot h_9(r)
  \label{a8}
\end{equation}
and also the tortoise coordinate $x$ by
\begin{equation}
  dx\ :=\ \frac{1}{G(r)}\cdot dr \quad.
  \label{a9}
\end{equation}
Since $G>0$ everywhere outside the horizon, the new coordinate $x$ is 
a regular, monotonic function of $r$ there. 
Thus, it may serve as an alternative radial coordinate for the
exterior of the black hole. Because of asymptotic flatness, $G$ comes close
to $1$ for large distances, and thus $x$ approaches infinity there.
Near the horizon, $G$ becomes zero, and hence $x$ tends to negative infinity. 
So $x$ parametrizes the whole exterior of the black hole.
Let us also look at the new dynamical variable $\Psi$. It is regular 
everywhere outside of the horizon, as $G$ and $R$ are regular there
and $R>0$. For $x\to+\infty$, $\Psi$ approaches zero faster than $1/x$,
but for $x\to-\infty$, $\Psi$ decreases even exponentially. Thus, $\Psi$
turns out to be a Fourier-transformable function.

With these new features, equation (\ref{a7}) now reads
\begin{equation} 
  -\partial_x^2 \Psi(x) +V_{eff}(x) \Psi(x)\ =\ 
  \omega^2 \Psi(x) \quad,\qquad\text{with}
  \label{a10}
\end{equation}
\begin{equation}
  V_{eff}(x)\ =\ \frac{G(x)}{R(x)^2}\bigl(L(L{+}1)-2\bigr)
  +R(x)\left(\partial_x^2 \frac{1}{R(x)}\right) \quad,
\end{equation}
which in fact is in Schr\"odinger-like form.  The effective potential belongs
to a special class of potentials; it is regular everywhere and tends to 
zero at both infinities.
For this kind of potential in one-dimensional Schr\"odinger problems, it
is well-known that there may exist a bound state only if the
potential is negative somewhere (see e.g \cite{Schwabl}). 
A bound state corresponds to a negative eigenvalue $\omega^2$, 
i.e. an unstable mode must exist. 
So a sufficient condition for stability is
\begin{equation}
  V_{eff}(x)\ \overset{!}{\ge}\ 0 \qquad\forall x\in[-\infty,+\infty] \quad.
  \label{a11}
\end{equation}

A further look at (\ref{a10}) reveals an angular
momentum barrier $\frac{G}{R^2}(L(L{+}1)-2)$, which grows 
monotonically with $L$, so that it is sufficient to perform the test on 
the mode with smallest value of $L$. 
The smallest non-trivial $L$ equals 2, and thus the final stability condition,
written in the old coordinate $r$, reads
\begin{equation} 
  V_{eff}(r)\ \overset{!}{\ge}\ 0 \qquad\forall r\in[h,+\infty]
  \quad,\qquad\text{with}
\end{equation}
\begin{equation}\label{Veffodd}
  V_{eff}(r)\ =\  
    4\frac{G(r)}{R(r)^2}+2\frac{G(r)^2{R'(r)}^2}{R(r)^2}
    -\frac{G(r)G'(r)R'(r)}{R(r)}-\frac{G(r)^2R''(r)}{R(r)} \;.
\end{equation}


\subsection{Even-Parity Sector}

Finally, we examine the even-parity perturbations. Again we first insert
(\ref{a1}), this time with the canonical even-parity ansatz (\ref{a2}),
into the equations of motion (\ref{eomcov}) and keep only
terms linear in $h_1$,$h_2$,$h_3$,$h_7$, and $\phi$. Again one
encounters derivatives of Legendre polynomials and reduces them
via their defining properties.
The resulting differential system is very complicated and difficult to
handle. It seems impossible to achieve a general Schr\"odinger-like 
form in the canonical gauge, as we just did in the odd-parity sector. 
We are searching for an alternative gauge and work on this subject 
is in progress.

Nevertheless, it is possible to achieve a Schr\"odinger-like form
of the differential equations in the special case $L=0$, which represents
monopole oscillations. For $L=0$ it is possible to gauge away two further
degrees of freedom and we choose
\begin{equation}
  h_1\ =\ 0\qquad\text{and}\qquad h_7\ =\ 0 \quad.
  \label{b1}
\end{equation}
This additional gauge-fixing results in an enormous simplification of the
differential equations. The (0,1) component of the gravitational 
perturbation equations is now easily solved for $h_3$,
\begin{equation}       
  h_3\ =\ -\frac{1}{2}\frac{R\Phi'}{GR'}\cdot\phi \quad.
  \label{b2}
\end{equation}       
Given this result, it is possible to solve the combination 
(2,2)+(2,3)$/\sin^2\vartheta$ of the components of the perturbation equations
for $h_2$,
\begin{equation}       
  h_2\ =\ \frac{i}{16\omega}\cdot\bigl(4\frac{RG\Phi''}{R'}
  +\frac{R^2G{\Phi'}^3}{{R'}^2}-4G\Phi'\bigr)\cdot\phi\
  +\ \frac{i}{4\omega}\frac{GR\Phi'}{R'}\cdot\phi' \quad.
  \label{b3}
\end{equation}
We have simplified the equation by using the static equations (\ref{eom})
to eliminate the function $V(\Phi)$.

Taken together, this allows us to decouple the scalar perturbation equation 
for $\phi$ (\ref{a2}) from the metric perturbations,
\begin{align}       
  0\ =&\ \Bigl(-\frac{1}{8}\frac{{\Phi'}^4R^2G}{{R'}^2}+\frac{3}{2}G{\Phi'}^2
  +\frac{\Phi'RG\Phi''}{R'}+\frac{1}{2}\frac{{\Phi'}^2RG'}{R'}
  -\partial_\Phi^2V(\Phi)+\frac{\omega^2}{G}\Bigr)\cdot\phi\nonumber\\
  &+\ \Bigl(2\frac{GR'}{R}+G'\Bigr)\cdot\phi'\ +\ G\cdot\phi''\quad.
  \label{b4}
\end{align}        
What is more, all other components of the perturbation equations 
turn out to be dependent or degenerate to the identity and, hence, give no
further information.
It is obvious that the second-order differential equation (\ref{b4})
determines the whole dynamics of the system, while 
(\ref{b2}) and (\ref{b3}) only express $h_2$ and $h_3$ in terms of
the dynamical variable $\phi$.

As in the odd-parity sector, it is possible to put the remaining 
equation (\ref{b4}) into a Schr\"odinger-like form by introducing
a new dynamical variable $\Psi$ and the tortoise coordinate $x$,
\begin{equation}
\Psi(r)\ :=\ R(r)\cdot\phi(r)\quad,\qquad\qquad 
dx\ :=\ \frac{1}{G(r)}\cdot dr\quad.
\end{equation}
The coordinate $x$ again parametrizes the whole exterior of the black hole,
and the new variable $\Psi$ is again regular everywhere. Because of
asymptotic flatness, $R\to r$ for large distances, and $\phi\to0$
like $1/r$ or faster. This means that $\Psi$ approaches
a constant at spatial infinity. The same behavior can be seen at the
horizon, where $x$ tends to negative infinity. $\Psi$ approaches
another constant here, so that $\Psi$ turns out to be a finite function 
and thus Fourier-transformable.
Expressed in the new variables, equation (\ref{b4}) now reads
\begin{equation}
  -\partial_x^2\cdot\Psi+V_{eff}\cdot\Psi\ =\ \omega^2\cdot\Psi
  \quad,\qquad{\text{with}}
  \label{b5}
\end{equation}  
\begin{equation}\label{Veffmono} 
V_{eff}\ =\ \frac{1}{8}\frac{{\Phi'}^4R^2G^2}{{R'}^2}-\frac{5}{4}G^2{\Phi'}^2
  -\frac{\Phi'RG^2\Phi''}{R'}-\frac{1}{2}\frac{{\Phi'}^2RGG'}{R'}
  +\frac{R'GG'}{R}+G\partial_\Phi^2V(\Phi)\;,
\end{equation}
and primes denote derivatives with respect to $r$. 
The effective potential $V_{eff}$ shows
a slightly different behavior at the two boundaries, namely
\begin{equation}
  \lim_{x\to-\infty}V_{eff}(x)\ =\ 0\qquad\qquad 
  \lim_{x\to+\infty}V_{eff}(x)\ =\ \partial_\Phi^2V(\Phi_0) \quad.
\end{equation}

There are two possibilities:
\begin{itemize}
  \item $\partial_\Phi^2V(\Phi_0)<0$: It is always possible
        to find a solution for $\Psi$ with negative eigenvalue $\omega^2$,
        thus the system is unstable.
  \item $\partial_\Phi^2V(\Phi_0)\ge0$: A positive semidefinite effective 
	potential is sufficient but not necessary to exclude bound states.
        If it is negative somewhere, a further investigation, 
	perhaps of numerical nature, must be performed.
\end{itemize}


\mysection{Stability}
The results of the previous two subsections suffice to examine the
stability properties of the static solutions constructed in Section~3. 
We shall apply the general criteria based on the effective potentials
(\ref{Veffodd}) and (\ref{Veffmono}) to each solution in turn.

\newpage

\subsection{Stability of the Coulombic Scalar Field Solution}
We first look at the odd-parity sector of the static solution
with power-like decaying hair (Section~3.1). The effective potential 
(\ref{Veffodd}) for this sector is shown in Figure \ref{figure10}. 
The effective potential vanishes exactly at the horizon 
(whose location moves with $Q$) with a positive slope,
and falls to zero exponentially at large distance.
It is positive everywhere exterior to the black hole, 
signalling stability.

For the even-parity monopole sector ($L=0$), the second of the
above two cases is relevant, since $\partial_\Phi^2V(0)=0$
for the Coulombic solution (\ref{Vcoul}).
Thus, the monopole sector is stable if the effective potential
never turns negative. The effective potential in this sector is shown in
Figures \ref{figure11} and \ref{figure12}. 
By inspection, the potential becomes negative for too large or too small 
values for the parameter $Q$, so that the system may become unstable there.
The upper stability bound for $Q$ can be found algebraically. Demanding 
positivity of the leading term at large distances, one finds
\begin{equation}
  Q^3\ <\ \frac{3}{2}\Tilde{M} \quad.
\end{equation}
The lower bound seems only accessible numerically. For $\Tilde{M}=1$,
one gets roughly $Q\approx-0.8$.

To summarize, the Coulombic scalar field solution (Section~3.1)
is completely stable against odd-parity perturbations, and is stable against
even-parity monopole perturbations at least for a certain range of $Q$.
The stability in the $L{>}0$ even-parity sector could not be decided here.

\figure[ht!]
\begin{center}\
\epsfbox[0 0 200 230]{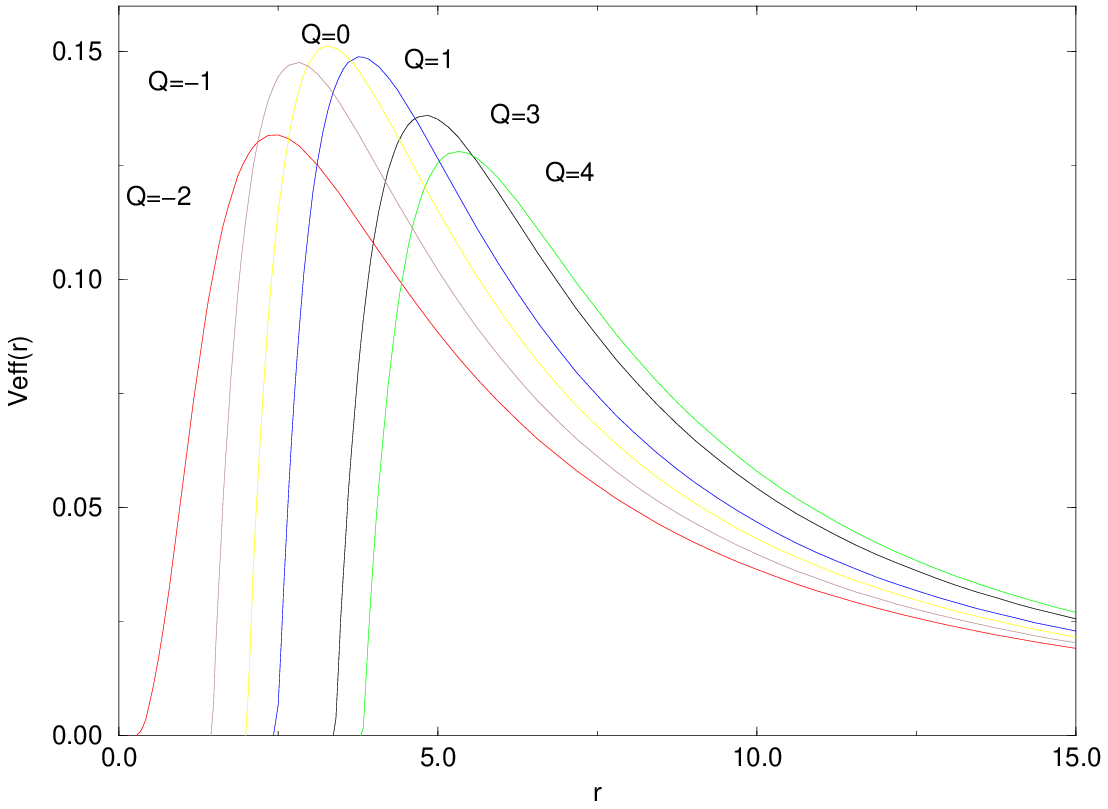}
\caption{$V_{eff}$ of the odd-parity sector for $\protect\Tilde{M}=1$
and several $Q>-2\protect\Tilde{M}$}
\label{figure10}
\end{center}
\endfigure

\newpage

\figure[ht!]
\begin{center}\
\epsfbox[0 0 200 230]{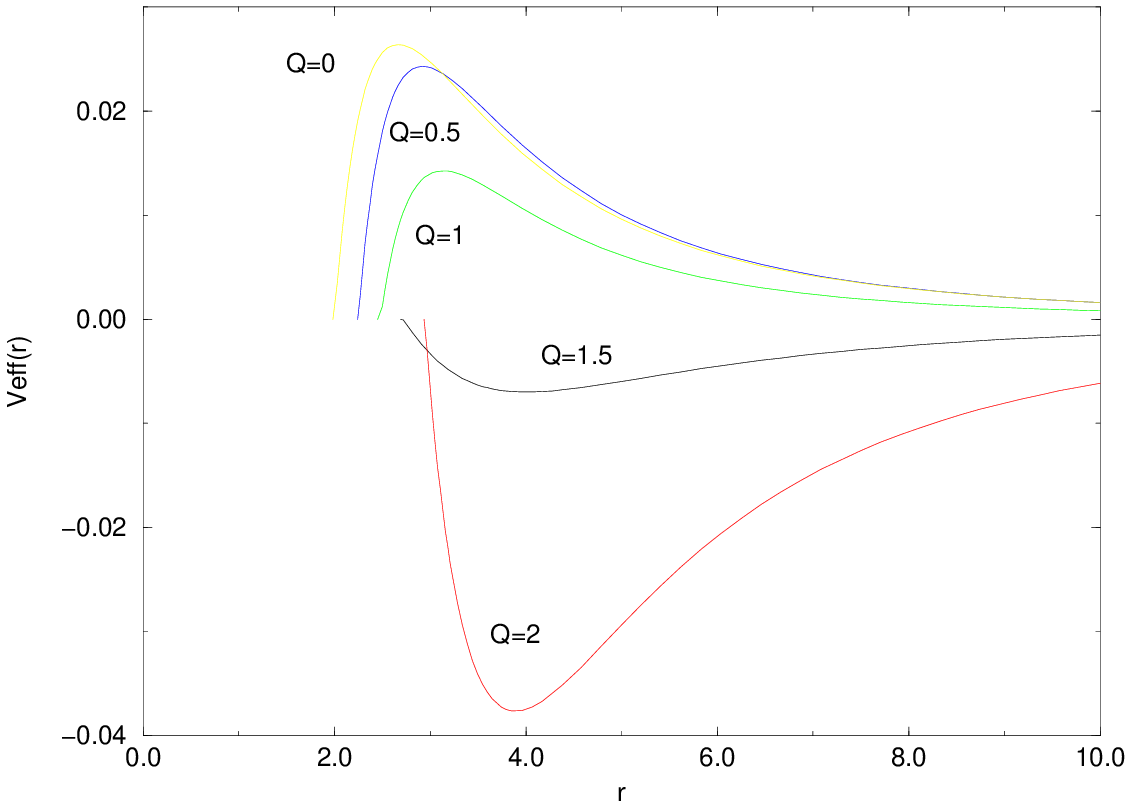}
\caption{$V_{eff}$ of the monopole sector with $\protect\Tilde{M}=1$
and positive $Q$}
\label{figure11}
\end{center}
\endfigure

\figure[ht!]
\begin{center}\
\epsfbox[0 0 200 230]{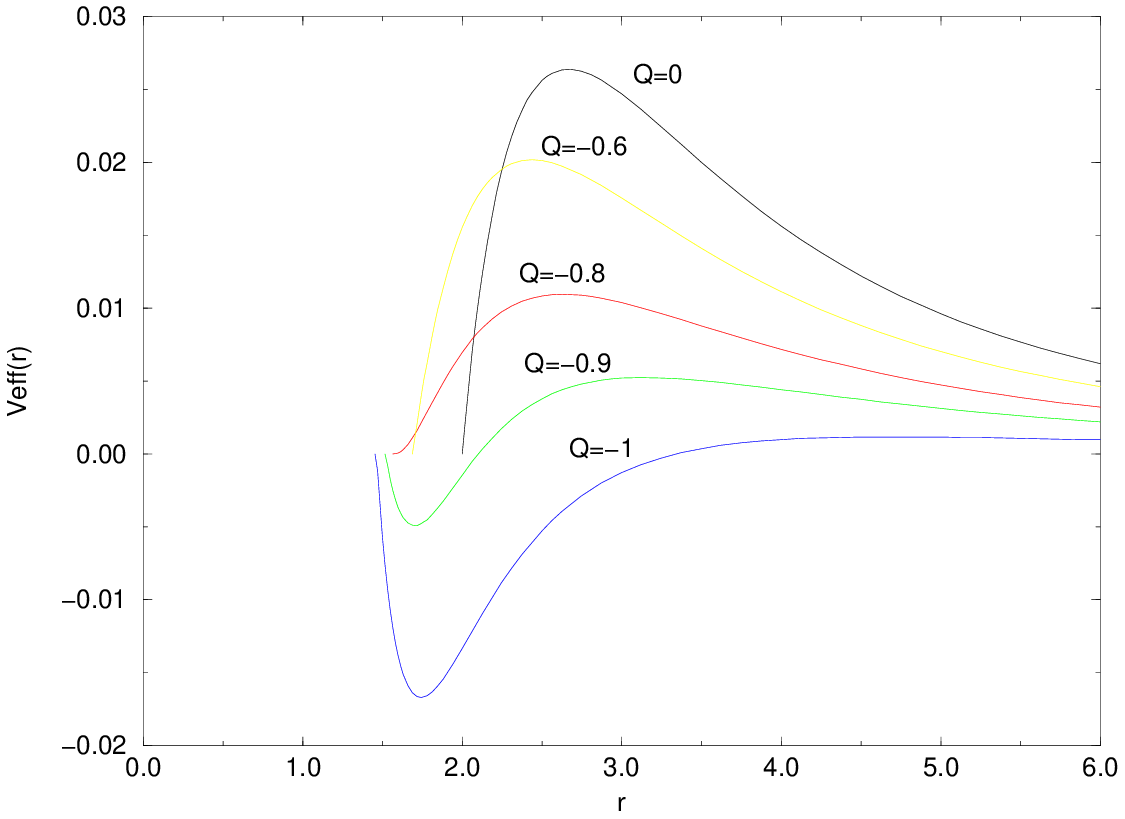}
\caption{$V_{eff}$ of the monopole sector with $\protect\Tilde{M}=1$
and negative $Q$}
\label{figure12}
\end{center}
\endfigure


\newpage

\subsection{Stability of the Exponential Scalar Field Solution}

As for the other static solution (Section 3.2), 
we first look again at the odd-parity sector.
The effective potential (\ref{Veffodd}) for this sector is shown in Figure
\ref{figure13}. It is surprising to observe that the effective potential 
changes only minimally with $q$, even though the static solution function $R$ 
varies dramatically within the same parameter domain 
(compare with Figure \ref{figure4}).
A zoom (Figure \ref{figure14}) reveals that the effective
potential does vary with $q$, as its zero always sits at the (moving) horizon.
The effective potential never turns negative, so the
system is stable against odd-parity perturbations.

Finally, we turn to the stability in the even-parity monopole sector.
This time, $\partial_\Phi^2V(0)=1>0$ from (\ref{Vexp}), so again we deal 
with the second case above.
This means that positivity of the effective potential is a
sufficient but not necessary criterion for stability.  The effective
potential for the monopole sector is shown in Figure \ref{figure15}.
As one can see there again, the effective potential depends only weakly on
the parameter $q$, while the static solution, especially $R$,
changes dramatically. A closer look at the horizon (Figure \ref{figure16})
shows the same feature as in the odd-parity sector.
With the effective potential being positive in the complete exterior of
the black hole, the system is stable against monopole perturbations.

To summarize, the exponentially decaying scalar field solution (Section~3.2)
is stable both against odd-parity and monopole perturbations. The answer to the
stability question for higher angular momentum in the even-parity sector 
still eludes us.

\newpage

\figure[ht!]
\begin{center}\
\epsfbox[0 0 200 230]{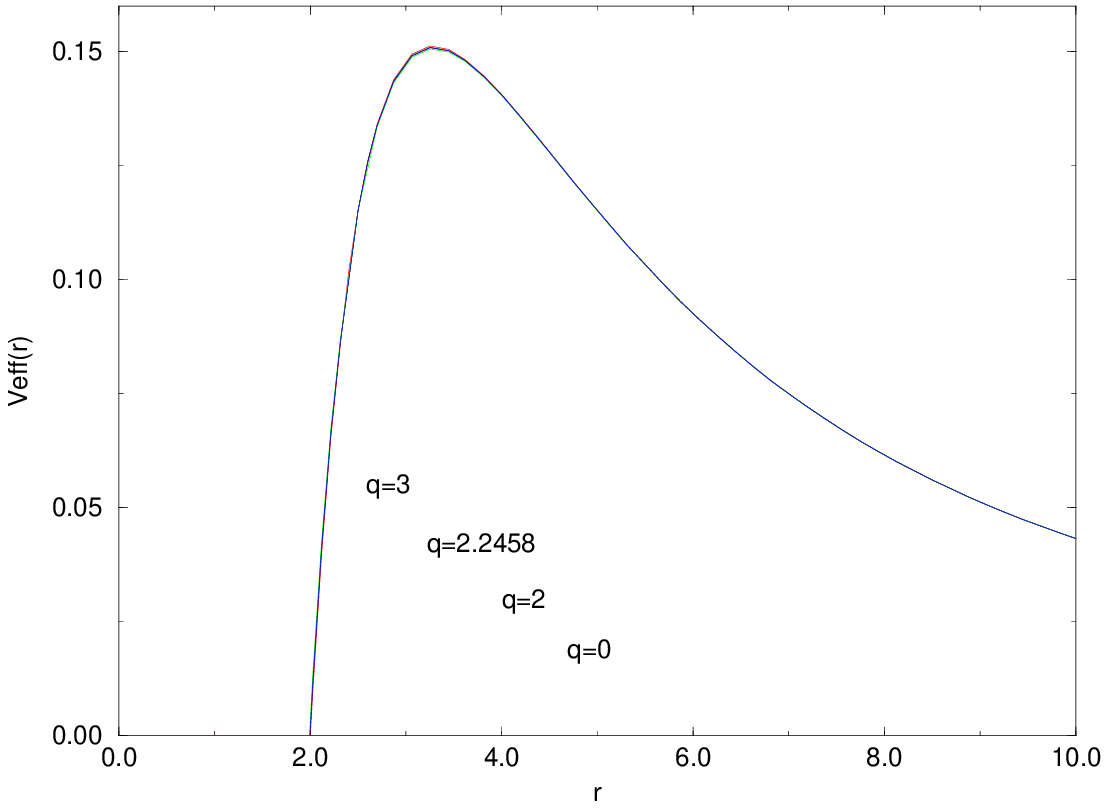}
\caption{$V_{eff}$ of the odd-parity sector with $M=1$ and several $q$}
\label{figure13}
\end{center}
\endfigure

\figure[ht!]
\begin{center}\
\epsfbox[0 0 200 230]{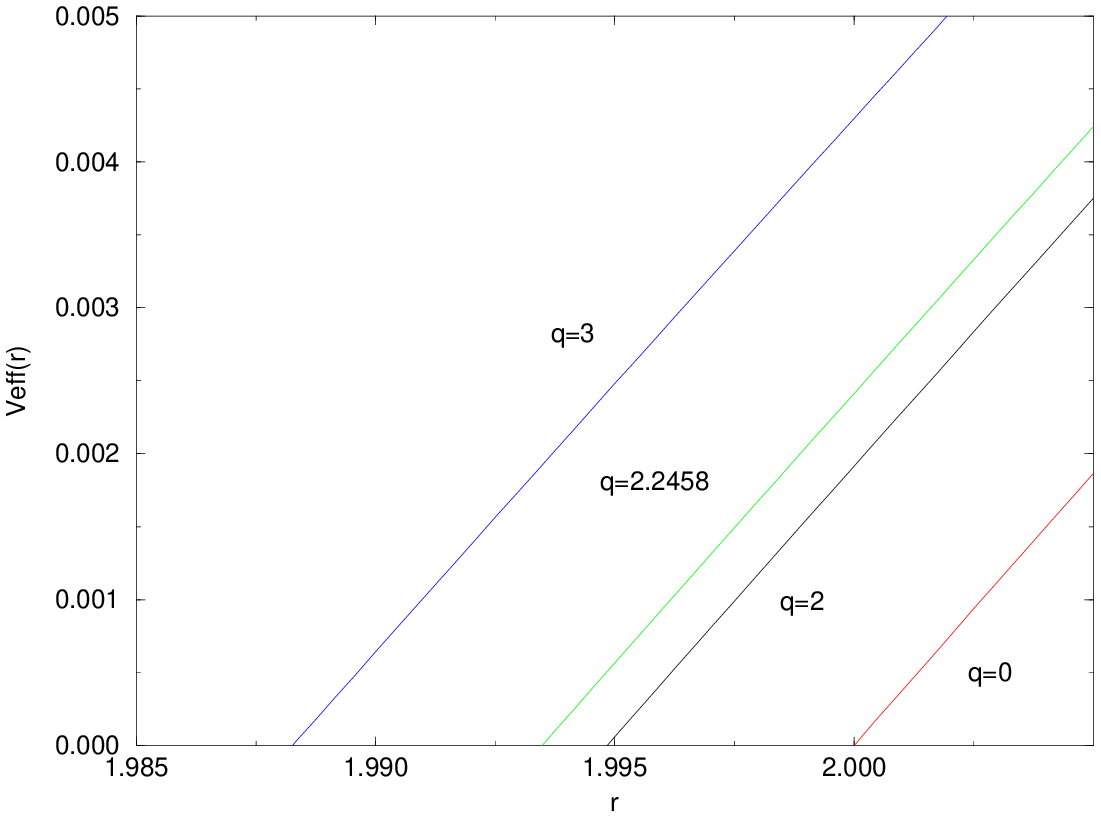}
\caption{Zoom of the odd-parity $V_{eff}$ around the horizon with $M=1$}
\label{figure14}
\end{center}
\endfigure

\newpage

\figure[ht!]
\begin{center}\
\epsfbox[0 0 200 230]{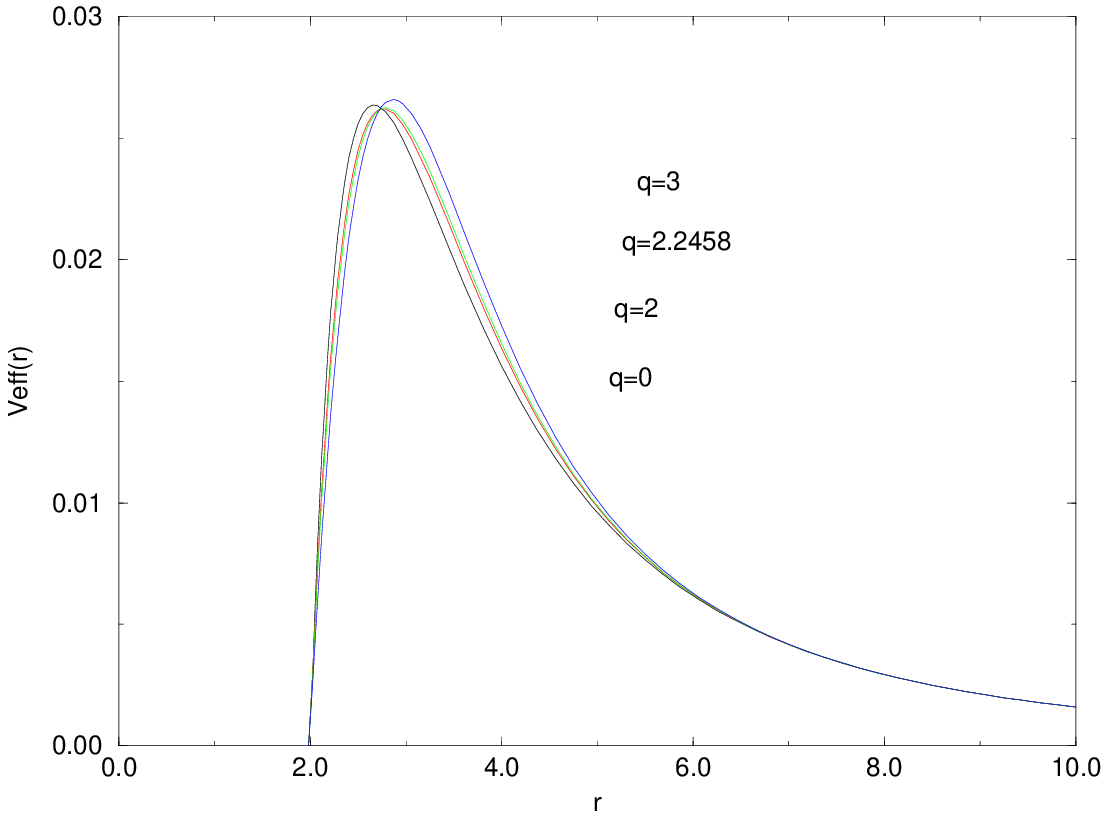}
\caption{$V_{eff}$ of the monopole sector with $M=1$ and several $q$}
\label{figure15}
\end{center}
\endfigure

\figure[ht!]
\begin{center}\
\epsfbox[0 0 200 200]{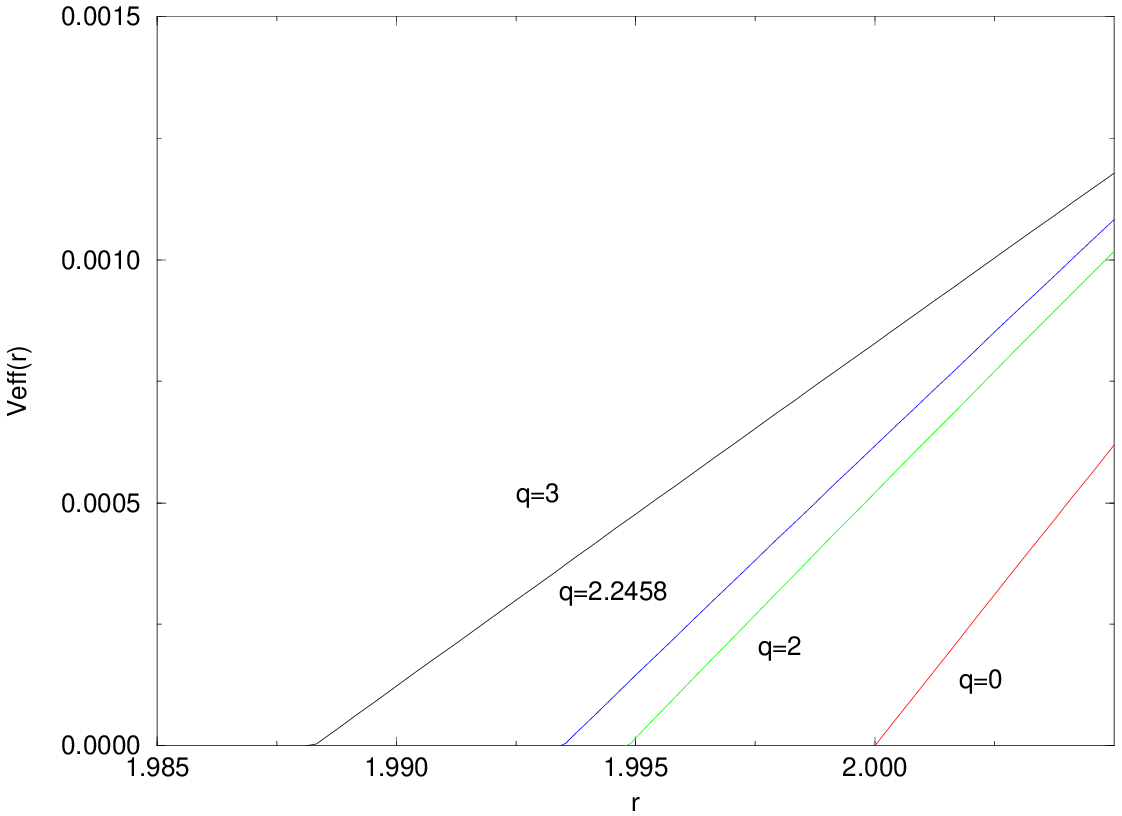}
\caption{Zoom of the monopole $V_{eff}$ around the horizon with $M=1$}
\label{figure16}
\end{center}
\endfigure


\newpage

\mysection{Conclusions}
We have demonstrated that it is possible to obtain asymptotically flat 
regular black-hole solutions of the static and isotropic Einstein--scalar 
field equations, if one gives up the dominant energy condition. 
These deformed Schwarzschild holes avoid the no-hair theorem by having 
partially negative scalar self-interaction potential.
In some cases, the metric and scalar configuration can be computed
analytically (we gave one example), and a generic feature seems to be
a mild non-analyticity of the potential at zero field.

The indefinite scalar potential does not imply instability however, at least
not against general odd-parity or monopole even-parity perturbations.
For these, we could derive sufficient stability conditions, which we then
tested on our examples, with positive results (in some parameter range).
Of course, this does not yet preclude potential instability against 
even-parity perturbations with higher angular momentum. Indeed, a numerical 
investigation seems to point in that direction~\cite{Marsa}.

There are a number of obvious extensions and open questions, in particular
\begin{itemize}
\item
Are there more analytic solutions?
\item
Can one add a cosmological constant?
\item
What effect does the scalar field have inside the horizon? 
How do our solutions look in Kruskal coordinates?
\item
With an additional gauge field, can one still find (partially) 
analytic solutions?
\item
May one generalize our solutions to the axisymmetric situation,
using the Ernst potential?
\item
Is there a minimal amount of violating the dominant/weak energy condition
for hair?
\end{itemize}
Some of these issues are under investigation, and we hope to report on them
in the near future.




\newpage

\begin{thebibliography}{99}
\bibitem{Bekenstein1}  J.D. Bekenstein,
                        {\it ``Black hole hair: twenty-five years after''},
                        gr-qc/9605059.
\bibitem{Heusler}      M. Heusler,
			{\it ``Black Hole Uniqueness Theorems''},\\
			Cambridge Lecture Notes in Physics 6
			(Cambridge U. Press, 1996).
\bibitem{Bekenstein2}  A.E. Mayo, J.D. Bekenstein,
			{\sl Phys. Rev.} {\bf D}54, 5059 (1996).
\bibitem{Bechmann}     O. Bechmann and O. Lechtenfeld,
                        {\sl Class. Quantum Grav.} 12, 1473 (1995).
\bibitem{Regge}        T. Regge and J.A. Wheeler, 
			{\sl Phys. Rev.} 108, 1063 (1957).
\bibitem{Vishveshwara} C.V. Vishveshwara, 
                       	{\sl Phys. Rev.} {\bf D}1, 2870 (1970);\\
                       C.V. Vishveshwara and L.A. Edelstein, 
			{\sl Phys. Rev.} {\bf D}1, 3544 (1970).
\bibitem{Zerilli}      F.J. Zerilli, 
			{\sl Phys. Rev. Lett.} 24, 737 (1970).
\bibitem{Buchdahl}     H. Buchdahl, 
			{\sl Phys. Rev.} 115, 1325 (1959).
\bibitem{Schwabl}      F. Schwabl, {\it ``Quantenmechanik''} 
			(Springer Verlag Berlin Heidelberg, 1990).
\bibitem{Marsa}	       R.L. Marsa, M.W. Choptuik,
			{\it ``Black hole--scalar field interactions in
			spherical symmetry''}, gr-qc/9607034;\\
		       R.L. Marsa, PhD Thesis, U. Texas at Austin (1995).
\end{thebibliography}
\end{document}